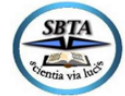

# BLOCKCHAIN-BASED TRAFFIC MANAGEMENT FOR ADVANCED AIR MOBILITY

Ítalo Romani de Oliveira, Thiago Matsumoto
Boeing Research & Technology Brazil

Euclides Carlos Pinto Neto
IEEE Member

**\* Corresponding author e-mail address**: italo.romanideoliveira@boeing.com

**PAPER ID: SIT132**

## ABSTRACT

The large public interest in Advanced Air Mobility (AAM) will soon lead to congested skies overhead cities, analogously to what happened with other transportation means, including commercial aviation. In the latter case, the combination of large distances and demanded number flights is such that a system with centralized control, with most of the traffic control decisions made by human operators, is safe. However, for AAM, it is expected a much higher demand for flights, because it will be used for people's daily commutes. Thus, higher automation levels will become a requirement for coordinating this traffic, because of its higher complexity, which might not be effectively managed by humans. The establishment of fixed air routes can abate complexity, however at the cost of limiting traffic capacity and decreasing flight efficiency. Another alternative is the use of a powerful central system based on Artificial Intelligence (AI) and / or other advanced algorithms, which would allow flexible trajectories and higher efficiency. However, such system would require concentrated investment, could contain Single-Points-of-Failure (SPoFs), would be a highly sought target of malicious attacks, and would be subject to periods of unavailability.

This work proposes a new technology that solves the problem of managing the high complexity of the AAM traffic with a secure distributed approach, without the need for a proprietary centralized automation system. This technology enables distributed airspace allocation management and conflict resolution by means of trusted shared data structures and associated smart contracts running on a blockchain ecosystem. This way, it greatly reduces the risk of system outages due to SPoFs, by allowing peer-to-peer conflict resolution, and being more resilient to failures in the ground communication infrastructure. Furthermore, it provides priority-based balancing mechanisms that help to regulate fairness among participants in the utilization of the airspace.

# 1. INTRODUCTION

## 1.1. ADVANCED AIR MOBILITY

Advanced Air Mobility (AAM) is an air transportation system concept that makes intensive use of disruptively new aircraft designs and flight technologies in order to enable alternative ways of moving goods and people in shorter times and more conveniently. It involves the so-called Urban Air Mobility (UAM), which uses highly automated aircraft operating the transport of passengers or cargo at lower altitudes within urban and suburban areas, and extended application cases, such as (FAA, 2022):

- Innovative Inter-city transportation;
- Public services;
- Private and Recreational uses.

## 1.2. TRAFFIC MANAGEMENT IN AAM

As AAM becomes more usual and popular, high volumes of traffic of the new air vehicles will arise. Currently there is no operating solution for high-capacity AAM traffic, despite the existence of several published Concept of Operations documents (FAA, 2020; EmbraerX; Atech; Harris, 2020; UK Air Mobility Consortium, 2022). These documents claim that the use of airspace structures in this environment can provide safety and efficiency benefits and therefore should be one of the fundaments in place. However, the obligation to follow circulation corridors and/or to be restricted to small cells of airspace decrease the efficiency of individual flights by requiring longer routes. This efficiency loss may be compensated by the benefits of avoiding gridlocks and chaotic traffic, but fixed corridors also create new bottlenecks. By the use of airspace structures, one is giving away efficiency and capacity in order to decrease complexity and increase predictability (Sunil, et al., 2016; Bauranov & Rakas, 2021).

On the other hand, it is possible to think of an airspace environment where trajectories are completely free and each aircraft keeps separation from others by its own means, which may not be safe when there are several aircraft close to each other, as demonstrated in (Romani de Oliveira, et al., 2021). The patent document (United States Patente Nº US10332405B2, 2014) describes a technical solution for a high traffic density of air vehicles, where each vehicle performs its own separation completely independently, without mechanisms to enable collaborative traffic management and reward fair use of the airspace. In spite of that, that solution still requires some centralized entities to provide authentication and maintain geo-fencing and spatial-fencing databases, which, in the absence of redundancy schemes, constitute Single-Points-of-Failure (SPoFs).

The papers (Romani de Oliveira, Matsumoto, Pinto Neto, & Yu, 2021; Romani de Oliveira, et al., 2021) present an in-depth comparison between centralized, de-centralized and collaborative traffic solutions, identifying the strengths and weaknesses of each one. The centralized solution is the most efficient and safe one in high-density scenarios, however has SPoFs by definition and requires a long-term fairness mechanism; the fully decentralized solution is robust and fair among equals but is inefficient and less safe in high density scenarios; and the collaborative solution brings a balance of these two opposite solutions, including the property that it is free of deadlocks, to which the strictly decentralized solution is not foolproof.

For a more comprehensive review of different airspace designs for AAM, considering various aspects as interaction with the city and society, one can refer to (Bauranov & Rakas, 2021).

## 1.3. BLOCKCHAIN TECHNOLOGY

Recurring to a known analogy, the Internet was developed exactly to make network communications more robust. Nobody controls the Internet, and it is extremely hard to bring it down in large scale, because each cluster of nodes is completely autonomous. This was intentional since the beginning, because one of the most important drivers in its creation was to survive to nuclear attacks. The decentralization principle of the Internet was so successful in transforming our lives that it enabled many associated collaborative tools of work and leisure, such as the wikis, social media, Git and many others. The novel Air Traffic Management solution proposed in this paper has a similar decentralization principle as that of the Internet, with the goal of achieving robustness and resiliency.

In the realm of online transactions and contracts management, the decentralization principle gave rise to the so-called Distributed Ledger Technology (DLT), which has the goal of ensuring validity of data blocks shared among many users without the need of a trusted third party. These data blocks can be used for transactions, data storage or for any other computational operation, shared among users that do not necessarily trust each other, with arbitrarily high levels of data integrity. A Blockchain is a type of DLT in which all validated data blocks are stored in a single line of time, each block referencing its predecessor and validated by a hash field, as illustrated in Figure 1.

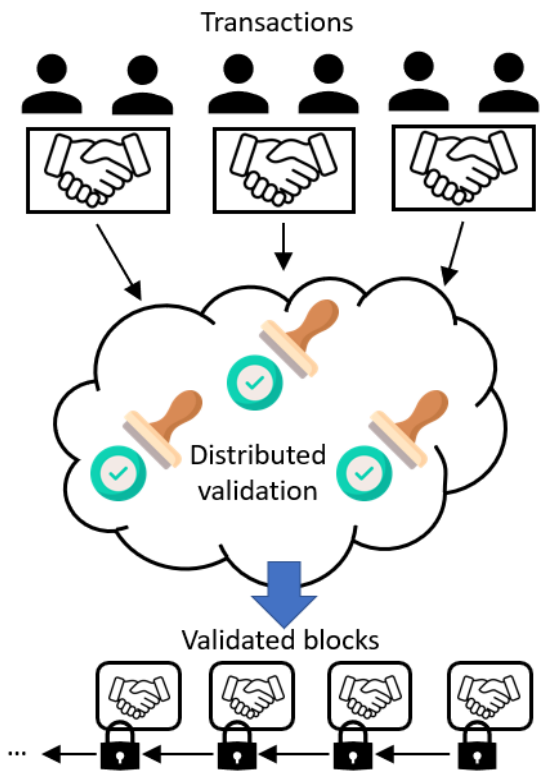

**Figure 1**: the Blockchain concept.

Blockchains became highly popular because they have the goal not only of being decentralized financial systems, but many of them are being used, appropriately or not, as means of wealth accumulation. The usefulness of Blockchain to the technical solution proposed here lies on various of its features: data integrity, irrefutability and the possibility of running complex logic inside it, that is, entire programs, also known as Smart Contracts. The consistency of these features is still evolving (Sultanik, et al., 2022), as a high degree of decentralization is difficult to attain in all aspects of the implementation, however we assume that a reliable blockchain solution will be available, being it either one of those existing today, or a new one to be developed.

As it will be shown in the next sections, complex logic is needed to be run inside the blockchain, in order to make collaborative or adversarial decisions with unambiguous and unescapable rules in the context of airspace allocation and conflict management.

## 2. THE PROPOSED SOLUTION FOR DISTRIBUTED AIR TRAFFIC MANAGEMENT

The main objectives of the solution here proposed is to provide high-capacity, reliable and safe automation for air traffic management, without the need for a trusted central authority that maintains complex software, infrastructure, and human operators. This section describes rules and logic that will run on a Blockchain to achieve those objectives, and some additional elements that are needed outside the Blockchain in order to complete the solution. For brevity, in the remainder of the paper we refer to this technical solution simply as "this system" or "system".

Here we describe the use case of a commercial or private air vehicle operator, but this use case system can be adapted to all types of airspace users, including public services operation such as medical, law enforcement, military, etc. The overall use-case logic of a flight is illustrated in **Figure 2** and explained in the next sections.

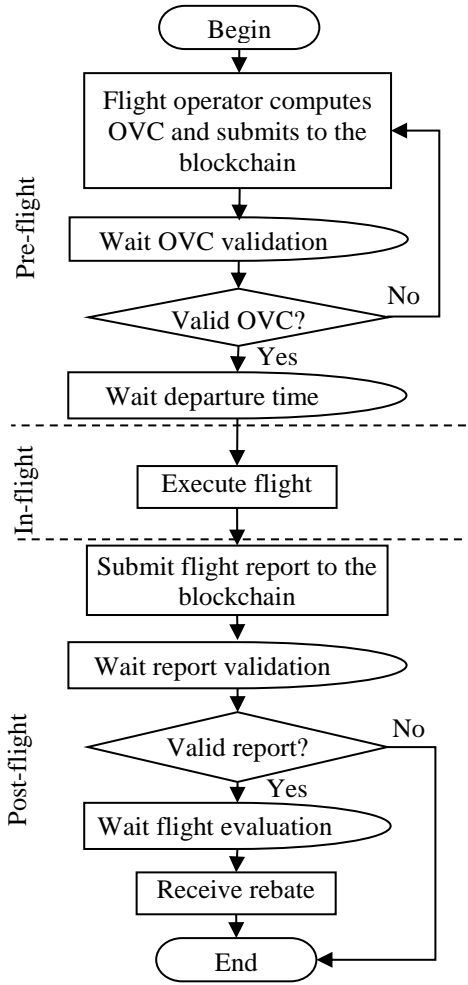

**Figure 2**: typical use case of a commercial or private flight operator

## 2.1. PRE-FLIGHT OPERATIONS

Let us consider the case of a vehicle operator $O_A$, which wants to fly from location $L_{A^0}$ to $L_{A^1}$. In order to deconflict its trajectory from others, the vehicle operator will compute the 4-dimensional volume which contains her vehicle's trajectory and additional safety buffers. This volume is named here Operating Volume Contract, or simply OVC (Hsieh, Sbai, Taylor, & Mitra, 2021). This OVC, here distinguished as $OVC_A$, will be thrown in the blockchain thread of unverified OVCs, and a token value will be deposited together with the tentative OVC as for maintenance of the blockchain and insurance for the case of not fulfilling the OVC. Another entity connected to this distributed system, here named $D_n$ ($n = 1, ..., N$), will search for conflicts between $OVC_A$ and other OVCs previously existing in the blockchain. If $D_n$ finds no conflicts with $OVC_A$, it will proceed to find the computational puzzle typical of a blockchain and provide a valid block, with corresponding hash value, which endorses $OVC_A$. This block may contain either only $OVC_A$ or bundle it with other verified OVCs. An abstract illustration of this idea is shown in **Figure 3**.

Any node in the blockchain (except the one to which $O_A$ is connected, by definition) might broadcast a validated block for $OVC_A$, without actually verifying its airspace safety. In order to prevent the bad consequences of this event, if any other node, let us name it as $D_m$, further finds a conflict with the already validated $OVC_A$, it will no longer forward the incorrect block, and generate a new block with a conflict demonstration, in a separate thread of the blockchain, the so-called conflict demonstration thread.

After some time elapsed, either the case of demonstrated conflict, or the case of absence of conflict, reaches a consensus among the blockchain nodes, and the corresponding block, let us name it as $B_k$, is accepted and stably incorporated to the blockchain, part of the deposit initially made by $O_A$, in the issuance of $OVC_A$, will be transferred to the owner of the node which first generated $B_k$, as a transaction fee to cover computation and networking costs. This incentive scheme works either if all nodes are equally verifying conflicts (analogously to the Proof-of-Work, or PoW, blockchain scheme), and only the first to generate $B_k$ gets rewarded, or a small number of randomly chosen nodes are designated to endorse or counter-proof the solicited OVC (analogously to the Proof-of-Stake, or PoS, blockchain scheme), and all of them get a reward. The latter case seems to be economically and environmentally more sustainable.

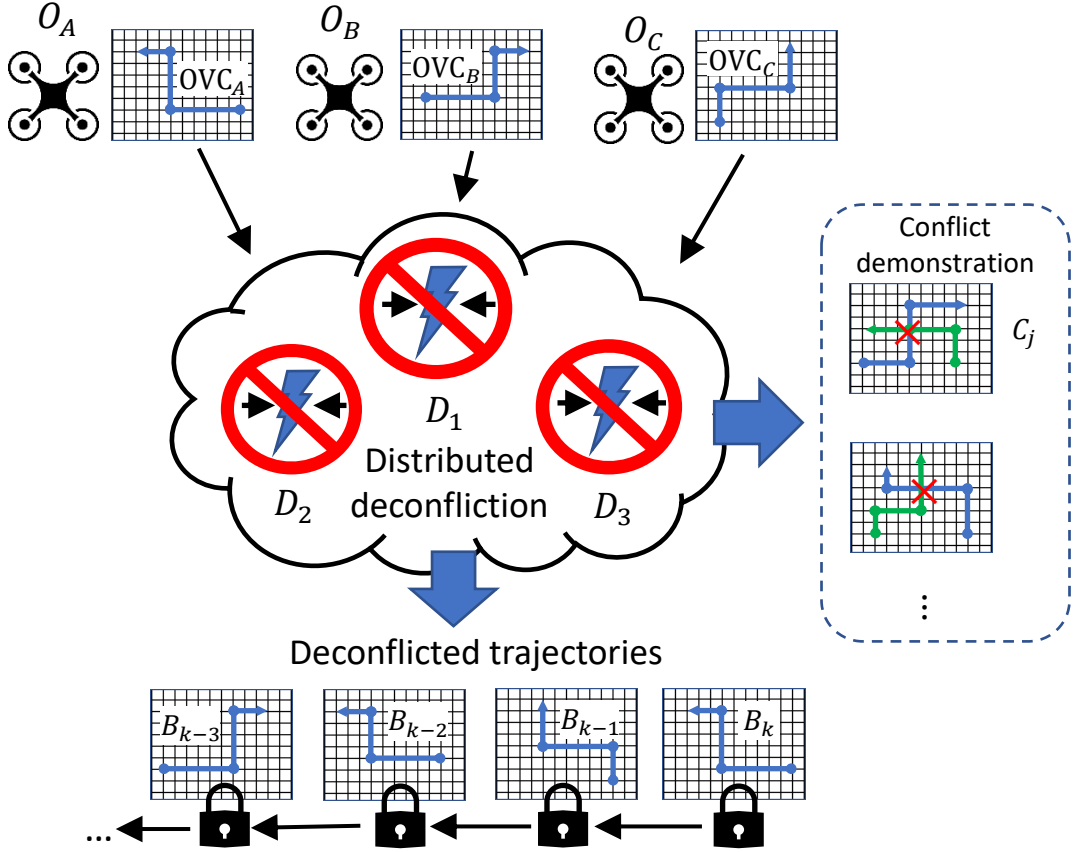

**Figure 3**: Blockchain-based validation of Operation Volume Contracts (OVC).

So, at any given moment, the blockchain thread of the planned OVCs would contain a number of OVCs planned by the various vehicle operators, with a guarantee that these OVCs do not interfere with each other and, at least in the planning context, the vehicles are safely separated from each other. The blockchain consensus scheme prevents that fraudulent users or operators disseminate OVCs that have not been validated by the consensus algorithm, unless one succeeds in performing the so-called 51% attack, which is practically impossible with the correct blockchain definitions and implementations.

This system can accommodate flights without a pre-defined trajectory, as it may happen for media coverage of events, certain types of observation flights, leisure flights, etc. In these cases, the operator requests a low-priority, non-exclusive OVC, which might be quite extensive. This volume is flagged as non-exclusive and is verified by other blockchain nodes, which check if the sum of simultaneous users of that portion of airspace does not exceed a certain limit (a capacity verification). The requested volume might include exclusive OVCs, for simplifying computations of 4D airspace volumes, however in that case the low-priority requesting user must not enter the exclusive OVC when it is actually in use.

The distribution and acquisition of token values to the users is out of the scope of the present paper, however, it might be controlled by the airspace authority, thus guaranteeing its authority role. The airspace authority may have the power to generate transactions to credit and withdraw token amounts from operators and participating nodes based on access rights and compliance to the rules of safe airspace use (e.g. to charge fines, etc.). However, most of the token transactions should be designed to incentivize efficient use of the airspace and fund the maintenance and technological progress of this system.

## 2.2. IN-FLIGHT OPERATIONS

Predictability is a strong support to flight efficiency. Thus, this solution allows mechanisms to incentivize predictability. On the one hand, air vehicle operations are subject to many uncertainty factors and should not be stifled by unreasonable requirements of predictability. One of the most important attractions of AAM is flexibility, which is enabled by this system, but in balance with predictability and safety. This is achieved by making that operators to recover part of the token amount that they deposited to issue a new flight, depending on an "assiduity" or "compliance" measure, which will evaluate how much the actual flight complies with the planned flight. This fundamental idea seems simple but, in practice, requires treatment of exception cases.

There are factors that are uncontrollable and/or unforeseeable by the flight operator, that may compromise a flight's compliance. For example, a vehicle is struck by a bird while in flight, having to do an emergency flight termination. While this is a rare occurrence, this case should not be regarded as a lack of assiduity to the OVC. But other more common occurrences are analogous, such as giving priority to a public service vehicle, with higher priority, or deviating from a blunder vehicle. Thus, after a flight terminates, the operator will issue a report of the flight, which will be the key for defining how much she will recover from the pre-flight deposit.

If an airspace conflict is detected while in flight, this system has definitions that help solving the conflict in a fair and efficient manner. Conflicts have three different categories:

1. Vehicles of different priority levels are involved in an encounter;
2. Some of the vehicles is out of its OVC;
3. Two or more vehicles with non-exclusive OVC are involved in an encounter.

In the first case, the vehicle with lower priority has to perform a deviation maneuver, while the vehicle with higher priority should not deviate. For example, a commercial vehicle has to deviate from a public service vehicle; or a vehicle with non-exclusive OVC has to deviate from a vehicle possessing an OVC and adhering to it.

In a second case, a vehicle requested an OVC (exclusive or non-exclusive) and is not adhering to it. In this case, it receives the lowest priority level and has to perform the deviation maneuvers from vehicles that are adhering to the OVC.

In the third case, the vehicles are in the same priority level, and a tiebreaker criterion is used, which is to compare the numeric value of hash obtained in the approval of its OVC. This criterion is also used in other cases where the vehicles are in the same priority level (for example, when both vehicles are deviating from the OVC).

This priority hierarchy is a fundamental aspect of this system, since it guarantees that conflicts are solved in realistically feasible amounts of time, a feature whose importance grows with the number of vehicles involved in a conflict: an aircraft has to perform a maneuver that takes into account only the other aircraft with lower priority; and avoids certain situations where repeated maneuvers could continue indefinitely (timeout situations).

## 2.3. POST-FLIGHT OPERATIONS

A flight report consists of a series of 4-dimensional points and, if deviations from the OVC occurred, justifications may be included. The report has to wait out a challenge period in order to obtain validation, according to Figure 4.

Any node in the network can initiate a challenge to this report and, in order to do that, has to deposit a token amount equal to that deposited by the issuer of the report. Once the challenge is initiated, it has a period to gather a committee with a minimum number of voters (e.g., 3); if not enough voters are gathered, the amount deposited by the challenger party will be returned to itself. On the contrary, if there are enough voters, the reward outcome would be as following.

If the challenging party wins, it will recover its deposit, and divide the amount deposited by the report issuer, after deducing other transaction fees, with the other voters. Otherwise, the challenging party will divide the amount deposited by the challenger with the other voters. This incentive scheme is aimed at promoting participation and truthfulness in a consensus-based verification of the fulfilment of the flight OVCs.

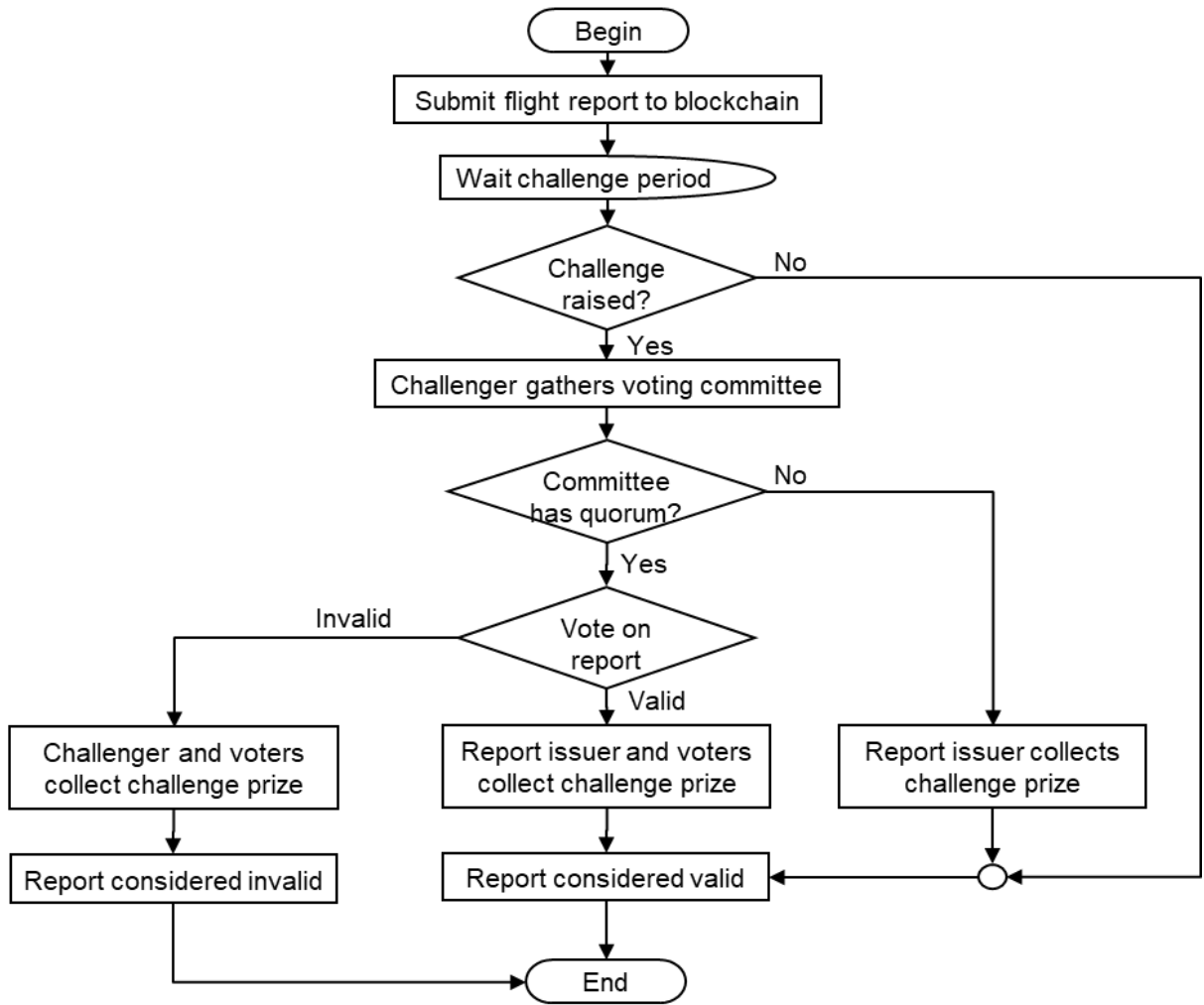

**Figure 4:** flight report validation process.

If no challenges to a report is issued during the challenging period, or if the report wins the challenges, it is incorporated as a valid report to the blockchain, and another consensus protocol is employed to provide its evaluation measure. In the case that the report is refused, there might be still a playoff request with a revised report. The evaluation measure calculation method must be uniform and universally accepted, and the consensus protocol should eliminate erroneous calculations and choose the prevalent value, as opposing to calculate a mean value. As in other consensus decisions, it may be either be via PoW or PoS, with corresponding distribution of transaction fees for the participating nodes, from the value deposited by the report issuer. The balance remaining from this deposit, after the various transaction fees, will be the base value to calculate its rebate. Good flight compliance evaluation measures will result in high rebates, while bad flight compliance evaluation measures fill result in low rebates. The balance that remains after all fees and rebates, from the initial deposit by the flight issuer, after all expiration deadlines, returns to the airspace authority, which will choose to redistribute it according to its policies of incentives and taxation.

## 2.4. PHYSICALLY TRACKING AIRCRAFT'S POSITIONS

Physical verification of vehicle location can be achieved by independent agents in various manners, but the one chosen for this system is by radio triangulation / multilateration, as illustrated by Figure 5.

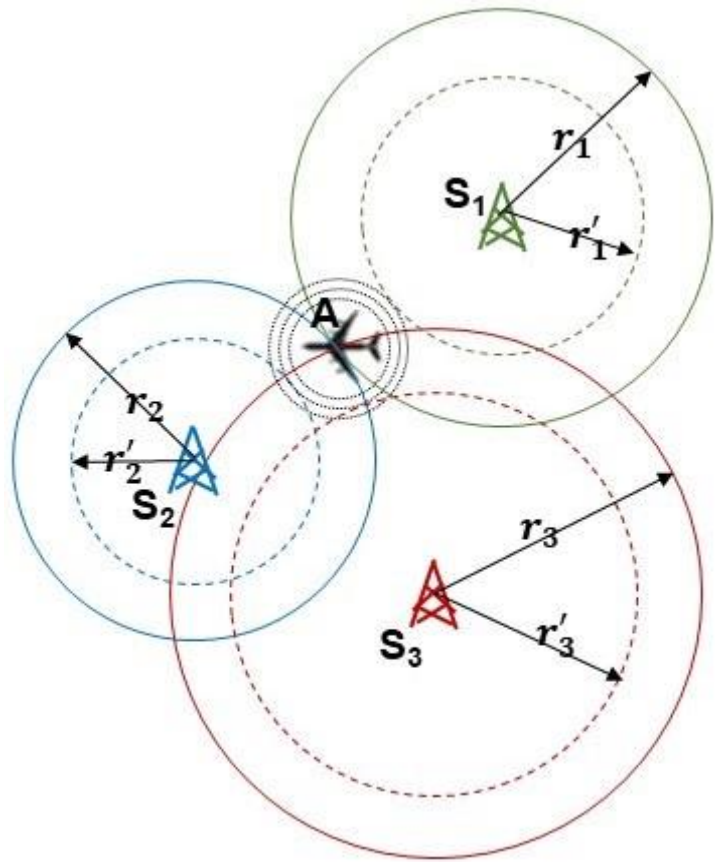

**Figure 5:** Multilateration of an aircraft's position via radio stations (source: Wikipedia).

Multilateration uses a principle that mirrors that of the Global Navigation Satellite System (GNSS), with position calculation performed outside the vehicle and belonging to the system composed by the external sensor stations ($S_1,…,S_n$). These stations may be fixed on ground, on satellites or even be dynamically composed by peer vehicles. Using records of GNSS signals would not be a reliable solution, since these signals can be spoofed (artificially generated). Apart from that, assuming that every aircraft can use a 4G or 5G cellular device, a report from the cellular telecom provider may be used for multilateration and attest vehicles' positions, however this might put too much power in the hands of the telecom provider, and incentivize it to collude with certain vehicle operators. Thus, in this solution the aircraft must be equipped with a radio device to broadcast an encoded message with its identity code, at least, in a non-directional signal, similarly to what is done with ADS-B for commercial and general aviation. This device can work in any frequency and modulation scheme that is appropriate to a low power consumption. Determining the spatial position (latitude, longitude, altitude) of an aircraft requires at least four passive stations, if the stations are synchronized, or six passive stations if they are not synchronized (Stefanski & Sadowski, 2018). If the aircraft transmits its altitude (measured by its own means), and this parameter is used as calculation input, one of these stations is no longer required. However, this system must be robust to erroneous or false altitudes reported by the aircraft (e.g. uncalibrated or faulty sensors), thus four or more stations are needed to determine an aircraft's position for each broadcast message transmitted. If fixed ground stations are used, it is expected that most of the stations in the vicinity of an aircraft will be positioned at similar altitudes, thus the accuracy of the altitude calculation via multilateration will not be very good with 4 stations only. However, by using filtering and machine learning methods, it is possible to gather data from an arbitrary number of stations and increase the accuracy of the position determination. Video processing techniques can also be used to help altitude and position determination. Satellite-based multilateration might be advantageous in altitude determination, although the ionospheric weather causes variability in the signal quality and, consequently, in the accuracy of position determination.

Passive receivers can be deployed and maintained by independent users, which can either triangulate the signals from various receivers, or offer their receivers' signals to pools, which in turn can triangulate the signals. The cost of this infrastructure would be funded by the blockchain transaction fees collected in the challenges and validity votes associated to a report and, if no challenges happen, in the final incorporation of the flight report to the blockchain. Various type of sensory data collected by the air vehicles also could be used to help determining other vehicles' positions.

Besides broadcasting its position and altitude, an aircraft participating in this system should broadcast its current OVC and corresponding hash value, provided by the blockchain. This information is used in conflict resolution, in the cases when airspace conflict happens with other aircraft while in flight.

A Blockchain-based solution for inviolable presence determination and registration of mobile entities was presented in (Leal, Pisani, & Endler, 2021), although it uses a variety of sensors and spatial conditions not comparable to the problem of position determination existing in the AAM context.

## 3. FINAL REMARKS

The system proposed must be implemented in a blockchain capable of smart contracts or in-block programming, either one of the existing ones, or one to be developed specifically with this purpose. It would be successful if a high proportion of the AAM vehicles in a certain region adhere to it. It is not foolproof against uncooperative vehicles (as any centralized system is), however it may be added with features to alert users about the uncooperative drones and register such occurrences with evidences and a certain level of consensus, in a way to discourage such practices.

Regardless of the implementation choices, if such choices are adherent to the definitions of this paper, a highly distributed and redundant system is obtained, where failures in one or a few nodes do not cause a breakdown of the whole system. The overall computational effort resulting from the summation of redundant computations and / or verifications will be probably higher than that of a centralized system with the same purpose, however the components of the system can be developed and verified collaboratively and achieve higher redundancy by having diverse systems and algorithms performing verification. Also, upgrades to the system can be seamless as its distributed nature allows that some nodes of the system are upgraded while others remain with their current versions, so to test updates gradually.

We also consider the possibility of combining some aspects of decentralization and blockchain with the presence of centralized USPs (Unmanned Air Traffic Management Service Providers), either by the USP being fully integrated with the decentralized services, or by the latter ones working as a secondary solution.

The overall concept of blockchain-based traffic management can be applied to surface (road and water) vehicles as well, however the features of these systems may not offer advantages for this concept as large as those envisioned for air vehicles.

## 4. ACKNOWLEDGEMENTS

We acknowledge the following authors for conceding their artwork used in our figures, obtained from the website [flaticon.com](flaticon.com):
- Uniconlabs for the handshake icon in Figure 1;
- Freepik for the stamp icon in Figure 1;
- Kosonicon for the drone icon in Figure 3.